\documentclass[aps,pra,floatfix,showpacs,noshowkeys,epsfig,graphics,natbib]{revtex4-1}
\usepackage{graphicx}
\usepackage{amsmath}
\usepackage{amsfonts}
\usepackage{amssymb}

\newcommand{\newc}{\newcommand}
\newc{\beq}    {\begin{equation}}
\newc{\eeq}    {\end{equation}}

\newc{\beqa}    {\begin{eqnarray}}
\newc{\eeqa}    {\end{eqnarray}}
\newc{\bs}    {\section}
\newc{\no}    {\\ \nonumber}

\newc{\st}    {\stackrel}

\begin{document}
\title{ Gravity as a Quantum Entanglement Force }
\author{Jae-Weon Lee}
%\email{scikid@gmail.com}
\affiliation{ Department of Renewable Energy, Jungwon
 University,  5 Dongburi, Goesan-eup, Goesan-gun, Chungbuk,
367-805, Korea\\
and \\ Asia Pacific Center for Theoretical Physic, Pohang University of Science and Technology,
 77 Cheongam-Ro, Nam-Gu, Pohang,
Gyeongbuk, 790-784, Korea
}
\author{Hyeong-Chan Kim}
%\email{hyeongchan@gmail.com}
\affiliation{School of Liberal Arts
and Sciences, Korea National University of Transportation, Chungju
380-702, Korea }

\author{Jungjai Lee}
\email{jjlee@daejin.ac.kr}
\affiliation{Department of Physics, Daejin University, Pocheon, Gyeonggi 487-711, Korea}

%\date{\today}

\begin{abstract}
We conjecture that  the total quantum entanglement of matter and
vacuum in the universe
   tends to increase with time,
 like  entropy, and that  an effective force
  is associated with this tendency.
We also suggest that gravity and dark energy are  types of quantum
entanglement forces, similar to Verlinde's entropic force,
and  give
holographic dark energy with  an equation of state
 comparable to  current observational data.
This connection between quantum entanglement and gravity
could give some new insights into the origins of gravity, dark energy,
 and  the arrow of time.
 \end{abstract}

%\pacs{98.80.Cq, 98.80.Es, 03.65.Ud}
\maketitle
\section{Introduction}

The recently proposed Verlinde's idea ~\cite{Verlinde:2010hp} linking the gravitational force to the entropic force
has attracted much attention
~\cite{Zhao:2010qw, Myung:2010jv, Liu:2010na,Tian:2010uy,Diego:2010ju,Vancea:2010vf,Konoplya:2010ak,Culetu:2010ua}.
He derived  Newton's equation and  Einstein's equation by using the relation.
Padmanabhan also proposed a similar idea~\cite{Padmanabhan:2009kr} by using the equipartition energy.
%However, Verlinde's proposal is based on  rather unusual assumptions such
%as the proportionality of the holographic entropy on the distance,
%and the holographic principle holding on equipotential surfaces.

In this paper, we conjecture that,  in general,
quantum entanglement of matter or the  vacuum in the universe
increases  like the entropy
and that  a new kind of
force (the `quantum entanglement force', henceforth), similar to the entropic force,
is associated with this tendency.
(This force is different from the `entanglement force' of polymer science.)
From this
perspective,  gravity and dark energy are  suggested to be
types of
 the quantum entanglement  force associated with
the increase in the entanglement,
similar to Verlinde's entropic force which is linked to the second law of thermodynamics.
Our model relies on the well-established quantum entanglement theory
and uses fewer  assumptions.

In a series of works~\cite{myDE,Kim:2007vx,Kim:2008re,Lee:2010bg,kias},
we have investigated the  quantum informational nature of gravity by utilizing
especially the quantum entanglement and  Landauer's principle.
Using the concepts, we  suggested   that dark energy  is related to the quantum
entanglement of the vacuum fluctuation~\cite{myDE}  at the cosmic
horizon~\cite{forget,Lee:2008vn} (or cosmic Hawking radiation~\cite{Lee:2008vn}) and that
the first law of black hole thermodynamics is derived from the second
law of thermodynamics~\cite{Kim:2007vx}.
Recently, we  also suggested~\cite{Lee:2010bg,kias} that the classical Einstein gravity could be derived
by considering quantum entanglement entropy and an information erasure at
Rindler horizons and Jacobson's idea linking the Einstein equation to
thermodynamics. All our results imply that gravity has something to do
with quantum information, especially quantum entanglement.
%Compared to Verlinde's proposal, our approach is based  on
% more ordinary quantum field theory.

 In Section II, we discuss the relation between entanglement and
 the holographic principle.
In Section III, we introduce the concept of the quantum entanglement force
and suggest that gravity is a kind of quantum entanglement force.
In Section VI, the predictions of our dark energy theory are compared with the recent observational data.
Section V contains discussions.

\section{Entanglement and holography}

In quantum information science, quantum entanglement
is a central concept
 and a precious resource allowing various types of
quantum information processing such as the quantum key
distribution.
Entanglement  is a  quantum nonlocal
correlation that cannot be prepared by  using local operations and
classical communication.
For pure states, the entanglement entropy $S_{Ent}$ is a good measure of entanglement.
For a bipartite system $AB$ described by a full density
matrix $\rho_{AB}$,
$S_{Ent}$  is the von Neumann entropy $S_{Ent}=-Tr(\rho_A ln \rho_A)$
for a reduced density matrix $\rho_A\equiv Tr_B
\rho_{AB}$ obtained by partial tracing  part B.
The partial tracing represents an ignorance of a subsystem.

For a typical example in quantum field theory, we consider a massless scalar field $\phi$  in a flat spacetime
with the Hamiltonian
 ~\cite{Srednicki}
\beq
H=\int d^3 x (|\nabla  \phi(x)|^2+ |\pi(x)|^2),
\eeq
where $\pi(x)$ is the  momentum of the field.
For a spherical region as shown in Fig. 1,
one can expand the field with spherical harmonics
on a discrete  radial coordinate with an UV-cutoff.
 An effective Hamiltonian for discretized field oscillators, $\phi_{lmj}$ ~\cite{Srednicki},
 contains  terms like
$\phi_{lmj}*\phi_{lm(j+1)}$, where $lmj$ are angular and radial indices.
These terms represent a nearest-neighbor interaction along the $r$ direction even at a causal horizon,
which can generate entanglement between the inside and the outside of the spherical surface.
One can find similar terms generating entanglement between two regions  for more general
spacetimes and fields.
The entanglement of the generic quantum field vacuum
 has  also been shown by using the Reeh-Schlieder theorem~\cite{reeh,werner}.

\begin{figure}[hbtp]
\includegraphics[width=0.32\textwidth]{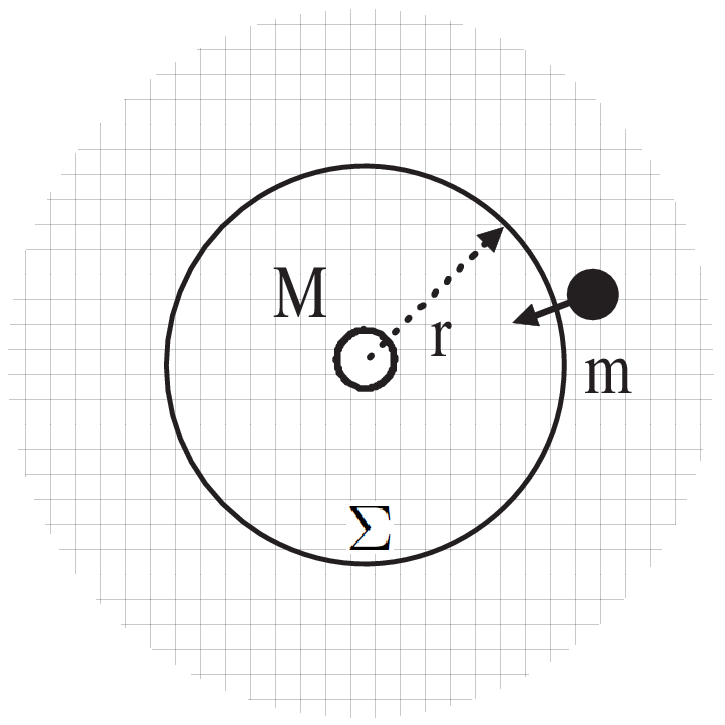}
\caption{ The space around a massive object with mass $M$
can be divided into two subspaces, the inside and the outside of an imaginary spherical
surface   $\Sigma$ with a radius $r$.
The surface $\Sigma$  has the entanglement entropy
$S_{ent}\propto r^2$ and entanglement energy
 $E_{ent}\equiv  \int_\Sigma T_{ent} dS_{ent}$.
If  a test particle with mass $m$ is present, it feels an effective attractive force
in the direction of the  increase in the entanglement between the inside and the outside of the surface.
}
\end{figure}
In general, the vacuum entanglement entropy of a spherical region with a radius $r$
 with quantum fields
can be expressed in the form of
\beq
 \label{Sent}
  S_{ent}=\frac{\beta r^2}{b^2},
  \eeq
   where $\beta$ is an $O(1)$ constant that
depends on the nature of the field and $b$ is the UV cutoff.
By performing numerical calculations on a sphere lattice,
Srednicki obtained  a value $\beta=0.3$ for the massless filed ~\cite{Srednicki}.
% $S_{Ent}$ has a form
%consistent with the holographic principle,  although it is derived
%from quantum field theory without using the principle.
A similar value was obtained in Ref. ~\cite{PhysRevD.52.4512,PhysRevLett.71.666} for the
entanglement entropy for a  massless scalar field in the
Friedmann universe.
More generally, we have to add the contributions from other fields with $\beta=\beta_j$ ~\cite{PhysRevD.52.4512}.
 If the j-th field has $N_j$  spin degrees of
freedom,
\beq
S_{ent}=\sum_j \beta_j N_j\frac{ r^2}{b^2}\equiv \frac{\alpha r^2}{l_P^2},
\eeq
where the Planck length $l_P=\sqrt{\hbar G/c^3}$.
If we choose $b=1/M_P$, where $M_P$ is the reduced Planck mass, then
\beq
\label{alpha}
\alpha=\frac{1}{8\pi}\sum_j \beta_j N_j.
\eeq
 Obtaining the value of $\alpha$ by using an explicit calculation in the future is important.
%where we put $1/8\pi$ for a later convenience.
The Bekenstein-Hawking entropy
\beq
S_{BH}=\frac{ c^3~A}{4G\hbar}
\eeq
was conjectured to
 saturate the information bound that a region of space with
 a surface area $A$ can contain~\cite{Bekenstein:1993dz}.
 If  $S_{ent}$ saturates this bound, i.e.,  $S_{ent}=S_{BH}$, then $\alpha=\pi$.

Why are we considering the quantum
 entanglement as an essential concept for  gravity?
First, interesting similarities exist between the holographic entropy
and the entanglement entropy of a given surface. Both are proportional to the area,
in general, and related to quantum nonlocality.
Second, when  a gravitational force exists,  a Rindler horizon always exists for some observers,
and it acts as an information barrier for the observers.
This can lead to ignorance of information beyond the horizons, and
the lost information can be naturally described by using the entanglement entropy ~\cite{Lee:2010bg}.
The spacetime should bend itself so that the increase in the entanglement entropy
compensates the lost information of matter. In  Ref.~\citenum{Lee:2010bg} we suggested
that the Einstein equation is  an
equation just describing this relation.
Third, if we use the entanglement entropy of quantum fields instead of the thermal
entropy of the holographic screen, we can understand the microstates of the
screen and, in principle, explicitly  calculate the relevant physical quantities by using  quantum field theory
in a curved spacetime. The microstates can
be thought of as  just quantum fields on the surface or their discretized oscillators.
On the other hand, if we identify the horizon entropy to be the ordinary thermal entropy of quantum fields,
we will encounter some problems.
The thermal entropy is a local quantity
that is incompatible with the holographic principle,
 and the thermal relaxation process  associated with the entropy
 may  be too slow to explain
the holograhic nature of a gravitational system with huge $r$.
 Finally, identifying the holographic entropy as the entanglement entropy could
explain why the derivations of the Einstein equation are involved with information
and, hence, quantum mechanics.
All these facts indicate that quantum mechanics and gravity have an intrinsic connection
and that the holographic principle itself has something to do with  quantum entanglement.

\section{Entanglement energy and entanglement force}

Separable (i.e., not entangled) states  are fragile in the sense that the states can be easily entangled with
environments surrounding the states.
A well-known example is the  Schr\"{o}dinger cat paradox~\cite{nielsen}.
In the paradox, no matter how well we separate a box  that contains the cat
from its environment,
we cannot fully block  the information leakage of the cat toward the environment outside the box.
Thus, even if we carefully prepare a superposition of the cat's state
$|dead ~cat \rangle  + |live ~cat \rangle $ ,
the state easily gets entangled with the environment to be
$|dead ~cat \rangle |env_0\rangle + |live ~cat \rangle |env_1\rangle$,
where $|env_0\rangle$ and $|env_1\rangle$ represent the corresponding states of the environment.
This entanglement between the cat and the environment induces a
decoherence of the cat's effective density matrix when we trace out the environment states.

Although this process is $reversible$ in principle,
practically, the reverse process is hardly observed on a macroscopic scale.
Ironically, this is one of the reasons observing a controllable
quantum entanglement
 in a laboratory
and  building a practical quantum computer using the entangled states is so difficult.
The quantum system of interest uncontrollably gets so easily  entangled with its environment
 and loses coherence within the system.
%(For a typical solid state system, the decoherence time scale is about femto seconds.)
Decoherence is also related to the emergence of the classical world from the quantum world~\cite{nielsen}.

 That the entanglement in the universe is increasing in general, considering that
 quantum evolution of a density matrix is described by  a unitary matrix $U$
as $\rho\rightarrow U^\dagger \rho U$, which is reversible, might seem strange.
 This paradoxical situation is very similar to the case with thermal entropy.
 Even though the Schr\"{o}dinger equation and the Einstein equation
 are time reversible, we see many time-irreversible phenomena in
 the macroscopic world, and the total entropy  does not  always decrease.
One way of handling this
`Loschmidt's paradox'
is to assume that the early universe had a very small entropy due to, for example, inflation.
Similarly, we can assume that the early universe began with a very small entanglement, too.
Thus, we can expect  the universe to have a strong tendency to increase the entanglement
among its constituents (matter, quantum fields, spacetime), as well as the entropy.
This might give us some new insights into the issue of the arrow of time.
The direction of time (i.e., the arrow of time) seems to be the direction in which
the  entanglement  increases. That is,
\beq
\frac{dS_{ent}}{dt}\ge 0
\eeq
for a sufficiently large macroscopic system and its environment.
In Ref. ~\citenum{Kim:2008re} we argued that the time evolution of the universe was related to the
expansion of the cosmic event horizon and its entanglement entropy.

Then, what is the relation between gravity and  entanglement?
In Ref. \citenum{myDE}, authors pointed out that a cosmic
horizon  had a kind of thermal
energy called entanglement energy related to $S_{ent}$,
\beq
\label{entenergy}
dE_{ent}\equiv  k_B T_{ent} dS_{ent},
\eeq
and suggested that it
was the origin of dark energy.
The above condition could be interpreted as extremization of the entanglement entropy with `heat' $dE_{ent}$.
This energy can be interpreted as the
effective energy obtained by
tracing out the Hilbert space describing the outside of the horizon.
It is also the energy of the vacuum fluctuation around the horizon.
In Ref. ~\citenum{Lee:2010bg} we pointed out that this energy was very similar to the equipartition energy
of the horizon~\cite{Padmanabhan:2009vy,Verlinde:2010hp}.
If this energy is a function of parameters $r_i$, one can define a generalized force
\beq
\label{entforce}
F_{ent,i}\equiv \frac{dE_{ent}}{dr_i}=k_B \partial_{r_i} (  \int T_{ent} dS_{ent})
=k_BT_{ent} \partial_{r_i} (  \int_\Sigma   dS_{ent}),
\eeq
which is similar to  the entropic force. We call this force  a `quantum entanglement force'.
At the last step, we assumed a surface integral on a
 $isothermal$  spherical surface (not equipotential) $\Sigma$.
%Thus, there are many parallels
% between our theory in ~\cite{Lee:2010bg} and Verlinde's theory
%in spite of some differences.

Now, we conjecture that the
quantum entanglement of matter and vacuum in the universe
   tends to increase over time
 like  entropy and that gravity is a kind of this quantum entanglement force,
similar to Verlinde's entropic force.
Below we will reinterpret  Verlinde's theory in terms of our entanglement theory.
To do this, we consider the situation in Fig. 1.
First, one can integrate Eq. (\ref{entenergy})
on the isothermal  spherical surface  $\Sigma$  with radius $r$ surrounding a mass $M$:
\beq
E_{ent}=\int_{\Sigma}  dE_{ent}= k_B T_{ent} \int_{\Sigma}  dS_{ent}=\frac{\hbar G M}{2\pi c r^2}
 \frac{\alpha r^2}{l_P^2}= \frac{\alpha M c^2}{2\pi} ,
\eeq
where we have used the Unruh temperature
\beq
T_U=\frac{\hbar a}{2\pi c k_B}=\frac{\hbar G M}{2\pi c k_B r^2}
\eeq
for $T_{ent}$, with  $a$ being the acceleration.
In Ref. ~\citenum{Lee:2010bg} we identified $T_U$ as the Rindler horizon temperature observed
by a test particle under the influence of the mass $M$.
For the holographic condition $E=Mc^2$ to hold on the surface,
 $\alpha$ should be $2\pi$, which exceeds the Bekenstein bound.
This discrepancy can be removed  by using the relation
$E=2k_B T S$ of Padmanabhan~\cite{Padmanabhan:2003pk}, which
seems to be valid when  an active gravitational mass exits. In that case,
we recover $\alpha=\pi$.

Now, we move on to the derivation of gravity. Consider a small test particle with mass $m$
 at a distance $r$ from the central object with mass $M$. This
will influence $S_{ent}$ of the spherical surface (Fig. 1). Let us denote this dependency as $S_{ent}(E_{ent},r)$.
Simply following Verlinde's approach, we express
the tendency to maximize the entanglement  by using the condition
\beq
 \frac{d S_{ent}(E_{ent}+ e^{V(r)}m,r)}{dr} ={\partial_r S_{ent}}+
 \frac{\partial S_{ent}}{\partial E_{ent}}\frac{\partial(e^{V(r)} m)}{\partial r} =0,
\eeq
where $e^{V(r)}$ represents the gravitational redshift with
some function $V(r)$.
This equation means that $S_{ent}$ increases as $r$ increases in such a way that the
newly-embodied mass $m$ at $r$ contributes $e^V m$ to $E_{ent}$.
Thus,
\beq
\partial_r S_{ent}=\frac{-\partial_r (e^{V(r)} m)}{k_B T_{ent}}.
\eeq

Then, this equation and Eq. (\ref{entenergy}) lead to \beq
F_{ent}=k_B   T_{ent} \partial_r \left( \int_\Sigma
dS_{ent}\right) =k_B   T_{ent} \partial_r S_{ent}= - m e^{V(r)}
\partial_r V(r). \eeq
In the weak gravity limit $GM \ll r$
$V\simeq -GM/r\ll 1$, $e^V\simeq 1$, and \beq F_{ent} \simeq - m
\partial_r V(r)=\frac{GMm}{r^2}, \eeq which is just  Newton's
gravity as Verlinde showed with the thermal entropy instead of
$S_{ent}$. Because we used the gravitational redshift for the
derivation, the appearance of Newton's gravity is not so
surprising. What we want to show  here is the relation between
gravity felt by the test particle and
 quantum entanglement of the whole system.
The test particle moves in a way that the total entanglement of the system maximizes.

\section{Entanglement and dark energy}

In Ref. \citenum{myDE}, we suggested  that a cosmic
causal horizon with a radius $R_h\sim O(H^{-1})$ had
a kind of thermal
energy $E_{h}\sim T_h S_h\propto R_h$, and that this energy was the dark energy. Here, $H$ is the Hubble
parameter, $T_h$ is the horizon temperature, and $S_h$ is its entropy.
To be specific, we considered the entanglement energy $E_{ent}$
associated with the cosmic event horizon for $E_{h}$.
 (Similar suggestions based on
the Verlinde's idea
~\cite{Li:2010cj,Zhang:2010hi,Wei:2010ww,Easson:2010av}
have appeared recently.)
Our theory can be easily extended with other cosmic horizons, such as apparent horions.

In this section, we will redo the calculation in  Ref. \citenum{myDE} except that we
integrate Eq. (\ref{entenergy}) on the horizon's surface instead of in the radial direction.
We will see that this gives an $E_{ent}$ that is a factor 2 smaller  than that Ref. \citenum{myDE}.

As before, by integrating $dE_{ent}$ on the surface
of the event horizon we obtain
 \beq
 \label{eent}
%E_{Ent}=\frac{\beta N_{dof}R_h}{\pi a^2},
E_{ent}=\int_{\Sigma}  dE_{ent}= k_B T_{ent} \int_{\Sigma}  dS_{ent}=\frac{\hbar c }{2\pi  R_h}
 \frac{\alpha R_h^2}{l_P^2}=\frac{ c^4 \alpha R_h }{2 \pi G} ,
\eeq
where we chose
 $T_{ent}= \hbar c /2\pi k_B R_h $, the Hawking-Gibbons temperature of the
horizon.
Using Eq. (\ref{entforce}), we find
 that this dark energy  corresponds to  a $constant$ quantum entanglement force
\beq
F_{ent}= \frac{dE_{ent}}{dR_h}=\frac{c^4 \alpha}{2\pi G},
\eeq
which makes the cosmological  horizon expand.
(A similar value for the entropic force was obtained independently
by Easson et al~\cite{Easson:2010av} using a surface action.)
Thus, we can say that dark energy is an effective force of the universe
associated with an increase of the quantum entanglement in the universe or in the
 area of the cosmic causal horizon.
Now, the entanglement
energy density of the cosmic event horizon is given by
\beq
\label{rho}
\rho_{\Lambda}=\frac{3 E_{ent}}{4 \pi R_h^3}=\frac{3 c^4 \alpha }{8 \pi^2 G R_h^2}
=\frac{3 c^3 \alpha  M_P^2}{\pi \hbar   R_h^2} \equiv \frac{3 d^2 c^3 M_P^2  }{\hbar  R_h^2 },
\eeq
which  has  the form  of
holographic dark energy~\cite{li-2004-603}.  From the above
equation,  we  immediately obtain a  formula for the constant:
 \beq
\label{d1}
 d=\sqrt{\frac{\alpha}{\pi}}.
 \eeq
If $S_{ent}$ saturates the Bekenstein bound, then $\alpha=\pi$; hence, $d=1$.
Before our works
the constant $d$ determining the equation of state $w_\Lambda$ of  the dark energy and
 the final fate of the universe,
 was obtained only by  observations  in Ref. \citenum{myDE}.
Theoretically, the value $d=1$ is  favored because it reproduces the de Sitter universe
when  the dark energy dominates.

\begin{figure}[htbp]
\includegraphics[width=0.4\textwidth]{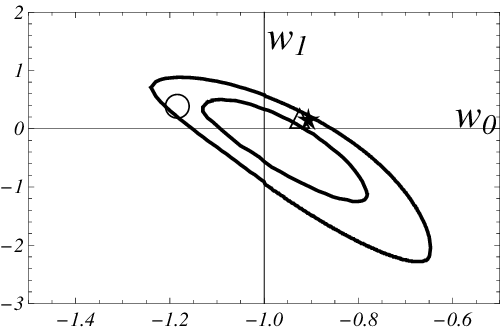}
\caption{Observational
constraint on
the dark energy equation of state
 $w_\Lambda (z) \simeq w_0+w_1 (1-R)$  from WMAP+BAO+$H_0$+SN. (The data are extracted from Fig. 13 in
 Ref. ~\citenum{Komatsu:2010fb}).
The contours show the $68\%$ and
the $95\%$ CLs, respectively.
 The star represents our theoretical prediction with $d=1$,
the circle is for the SM $(d=0.67)$ and the triangle is  for the MSSM $(d=0.962)$.
 \label{Fig1} }
\end{figure}

One can compare  predictions of our theory directly with current observational data.
The equation of state for
holographic dark energy  is given by  ~\cite{li-2004-603,1475-7516-2004-08-013}
\beq
\label{omega}
w_0 =-\frac{1}{3} \left(1+\frac{2\sqrt{\Omega^0_\Lambda} }{d}\right),
\eeq
and
its  change rate at  the present is ~\cite{li-2004-603,1475-7516-2004-08-006}
\beq
\label{omega3}
w_1
 =
\frac{\sqrt{\Omega^0_\Lambda} \left( 1 - \Omega^0_\Lambda  \right)}{3d} \left(1+\frac{2\sqrt{\Omega^0_\Lambda} }{d}\right),
\eeq
where  $z$ is the redshift parameter, $\Omega_\Lambda^0$ is the present  value of
the density parameter
for  dark energy,
 $w_\Lambda (z) \simeq w_0+w_1 (1-R)$, and $R$ is the scale factor of the universe.
 For $\Omega^0_\Lambda= 0.73$ and $d=1$,
 these equations give $w_0=-0.903$ and $w_1=0.208$.
 These theoretical values for $\omega_\Lambda$ are comparable to the
 current observational data.
 Although the cosmological constant is most favored by observations,
 a large range of values is still allowed by the data for the time-dependent $w_\Lambda$.
    For example, the combination of
 WMAP (Wilkinson Microwave Anisotropy Probe)  7-year data,
 the  baryon acoustic oscillation (BAO), Type Ia supernovae (SN), and the Hubble constant ($H_0$) data
  yields $\omega_0 = -0.93\pm 0.13$
  and  $w_1 = -0.41^{+0.72}_{-0.71}$ at the $68\%~ CL$~\cite{Komatsu:2010fb}.
(Note that the observational uncertainty  for $w_1$ is still large and  that  $w_1$ is  $0.11\pm 0.7$ in
the WMAP-5 year data~\cite{Komatsu:2008hk}.)

Alternatively, we can rely on the quantum field theory to avoid the use of the Bekenstein bound.
If we use the approximation $\beta_j \simeq 0.3$ for all $j$
and  use $\sum_j N_{j}=118$ of the standard model (SM) of particle physics in Eq. (\ref{alpha}),
  we obtain $d=0.67$, $w_0 \simeq  -1.18$ and  $w_1 \simeq 0.407$.
  For
  the minimal supersymmetric standard model (MSSM), $\sum_j N_{j}=244$,
   $d= 0.962$,  $\omega_0\simeq -0.925$
and $\omega_1\simeq 0.22$.
Thus, our theory with quantum field theory
is still in  good agreement with the observational data
and favors supersymmetric theories over non-supersymmetric ones.
(Recall that the $d$ values in this paper are half of those in   Ref. ~\citenum{myDE}.)

\section{Discussion}

 Understanding
the entropic origin of gravity is important.
In this work, we have tried to
reconcile Verlinde's theory with our theory based on quantum information.
 Many similarities  exists between the two theories.
If we identify the horizon entropy as the entanglement entropy and
the equipartition energy as the entanglement energy,
we can give the theory a better foundation.

We conjectured that the total quantum entanglement of matter and
fields in the universe
   tends to increase over time and that  an effective force is associated with this tendency.
   This force might be very general in the nature, and we expect
   that  one can measure this force in a quantum information experiment
   using quantum optics or solid-state quantum devices.

We also suggest that dark energy and, more fundamentally, gravity itself
are  quantum entanglement forces
similar to Verlinde's entropic force.
If the entanglement entropy of the universe saturates the Bekenstein bound,
this gives the holographic dark energy with  an equation of state
 comparable to the  current observational data.
 Our quantum informational interpretation of gravity may provide some new insights into  the natures of gravity, dark energy,
  and  the arrow of  time.

\section*{ACKNOWLEDGEMENT}
This work was supported by the Daejin University Research Grants 2015.

\end{document}